\documentclass[
 reprint,
 amsmath,
 amssymb,
 pra
]{revtex4-2}

\usepackage{graphicx}
\usepackage{dcolumn}
\usepackage{bm}
\usepackage[mathlines]{lineno}

\begin{document}

\preprint{aps/123-QED}

\title{Phase-space interpretation of spatial stationarity for coherence
holography}

\author{Rishabh Kumar Bhalavi and Rakesh Kumar Singh}
\email{rishabhkr.bhalavi.phy16@itbhu.ac.in\\ krakeshsingh.phy@iitbhu.ac.in}
\affiliation{Department of Physics, Indian Institute of Technology (BHU),
Varanasi, 221005, Uttar Pradesh, India}

\date{\today}

\begin{abstract}
We extend the wide-sense spatial stationarity concept of coherence
holography in the regime of phase-space using the wigner distribution
function. We focus mainly on the incoherent light source and the Fourier
and Fresnel propagation kernels for the optical-field transformation
rule (input-output relation) and derive the same analogy in WDF. We
further show that in phase-space the WDF obtained from the ensemble-averaged
and space-averaged coherence functions are the same. Finally, we interpret
behaviour of these results through numerical simulations.
\end{abstract}

\maketitle

\section{introduction}
In 1932, E. Wigner introduced the Wigner distribution function (WDF) to analyse quantum mechanical systems in phase-space \cite{wigner1932phys}. Later, Ville \cite{ville2orie,qian1999joint} and Bastiaans \cite{bastiaans1978wigner,bastiaans1979transport,bastiaans1979wigner} introduced this concept in Signal processing and Fourier optics, respectively. Since then, it has found numerous applications as being an excellent tool for the phase-space analysis with many properties \cite{bastiaans1980wigner,torre2005linear}. However, any of these property's applicability depends on the analysis being done. For example, being a bilinear function, it produces undesirable cross-WDF terms for some Holography and Signal processing applications \cite{situ2007holography,khan2011cross}.

In the past, there have been several publications on phase space interpretation of Holography. For example, Lohmann et al. proposed a Wigner chart method to examine the hologram recording materials and detectors' storage capacity, which is an easy way to determine recording trade-off for various Hologram recording schemes \cite{lohmann2002holography}. Likewise, Oh and Barbastath presented a way to obtain WDF of the volume hologram. To produce the result, they first got the WDF of the 4f imager and then used the WDF shearing properties \cite{oh2009wigner}. Next, Kim, Hwi, et al. proposed an optical sectioning method for optical sectioning holography. They showed that the focused and defocused information at the focal plane are separable in phase-space (WDF); hence, a filter mechanism can be built \cite{kim2008optical}. Likewise, one may find examples of varied applications of the WDF in Holography.

Holography is a widely used principle in optical and non-optical applications. It was first introduced by Gabor in 1948 as a microscopic principle to record complete information (phase and amplitude) of an object \cite{gabor1948new}. Later, Leith and Upatneiks modified this principle by introducing the off-set reference wave \cite{leith1962reconstructed,leith1963wavefront}. Since then, many conventional and non-conventional Holography techniques have been reported. One such non-conventional holography technique is the Coherence Holography proposed by Takeda et al. In this technique, the hologram is recorded with coherent light and reconstructed with incoherent light. The resulting image, upon reconstruction, is the distribution of spatial coherence function, obtained from the space average instead of the ensemble average. To further elaborate, we deal with spatial statistics of an optical field rather than its temporal statistics. If a statistical field is temporally stationary and ergodic, then the ensemble average is replaced by the time average. Usually, the coherence function is obtained from the ensemble average or time average, which is applicable for many practical cases. Nonetheless, modern advances in optics such as time-frozen optical fields can not be dealt with temporal statistics and motivates us to devise other methods as well. This is where the spatial statistics comes into the picture. When scattered from a diffusive material, such fields give a good insight into the object when dealt with spatial statistics. Although not much has been explored into the spatial statistical approach, the coherence holography elaborates an excellent use of it by replacing the ensemble average by the space average provided the wide-sense spatial staionarity condition is satisfied \cite{takeda2005coherence,takeda2013spatial,takeda2014spatial,o2003introduction}. In this letter, our attempt is to interpret this condition in phase-space and to supplement the underlying theory of coherence holography. Our primary focus here is the reconstruction with the Fourier and Fresnel kernels and the source obtained from the hologram illuminated with incoherent light. Now, as we proceed section-wise, we will elaborate more onto each concept.

Sec. 2. has three subsections A, B, and C. In A, we derive the input-output relation of the WDF for partially coherent light and introduce the double WDF concept. In B, we present the conditions for the validity of wide-sense spatial stationarity of the output coherence function. Then, we derive the Wigner counterparts of the input coherence function (from the source) and the Fourier and Fresnel kernels. Finally, in C., we present an alternative way to obtain the output WDFs from the spatial coherence functions. Next in sec. 3., we demonstrate simulated results and infer the behaviour of output WDFs and their relation to the input WDFs. Finally, in sec. 4., we conclude with some remarks on our result. 

\section{theory}

\subsection{Input-ouput relation}

\begin{figure}[h]
\includegraphics[scale=0.36]{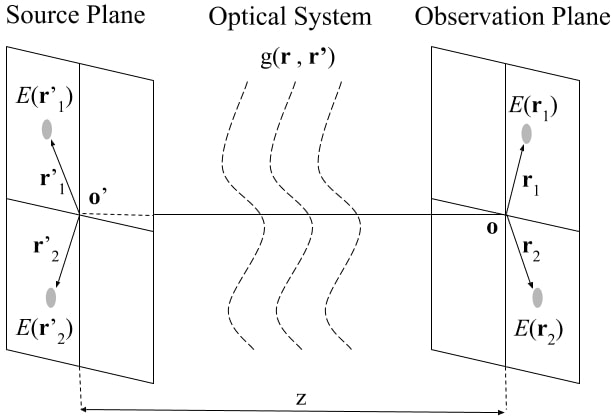}
\caption{Geometry of source plane, optical system, and observation plane.}
\end{figure}

Let us consider an input-output relation for an optical field, which
is coherent and quasi-monochromatic in nature.

\begin{equation}
E\left(\boldsymbol{r}\right)=\int g\left(\boldsymbol{r},\boldsymbol{r}'\right)E\left(\boldsymbol{r}'\right)dr'
\end{equation}

As shown in fig. 1, $E\left(\boldsymbol{r}\right)$ and $E\left(\boldsymbol{r}'\right)$
are the optical fields at the source $\left(\boldsymbol{r}'\right)$
and observation $\left(\boldsymbol{r}\right)$ planes, respectively.
The function $g\left(\boldsymbol{r}',\boldsymbol{r}\right)$ is the
line-spread function or green's function, describing the impulse-response
relation of the system. Also, the integration is taken over the entire
source plane unless otherwise stated. Now, if the light is partially
coherent, the field becomes a stochastic process, and its transformation
is studied in terms of coherence function or two-point correlation
function. The coherence function is defined as $\Gamma\left(\boldsymbol{r}_{1},\boldsymbol{r}_{2}\right)=\left\langle E^{*}\left(\boldsymbol{r}_{1}\right)E\left(\boldsymbol{r}_{2}\right)\right\rangle $,
where $E\left(\boldsymbol{r}\right)$ is the field and $\left\langle ...\right\rangle $
represents the ensemble average. One may easily derive the input-output
relation in terms of coherence function by using Eq. $(1)$ for $E\left(\boldsymbol{r}\right)$.
This input-output relation has an analogy in terms of WDF as well \cite{torre2005linear}. 

For that, we first consider the WDF definition of the input coherence
function (at the source plane),

\begin{equation}
\mathcal{W}\left(\boldsymbol{q}',\boldsymbol{k}'\right)=\frac{1}{2\pi}\int\Gamma\left(\boldsymbol{q}'-\frac{\boldsymbol{p}'}{2},\boldsymbol{q}'+\frac{\boldsymbol{p}'}{2}\right)\exp\left\{ -i\boldsymbol{k}'\cdot\boldsymbol{p}'\right\} d\boldsymbol{p}'
\end{equation}

It also implies the following \emph{inverse} WDF relation

\begin{equation}
\Gamma\left(\boldsymbol{q}'-\frac{\boldsymbol{p}'}{2},\boldsymbol{q}'+\frac{\boldsymbol{p}'}{2}\right)=\int\mathcal{W}\left(\boldsymbol{q}',\boldsymbol{k}'\right)\exp\left\{ i\boldsymbol{k}'\cdot\boldsymbol{p}'\right\} d\boldsymbol{k}'
\end{equation}

Next, we consider the \emph{double} WDF definition, which is a basic
definition for our analysis.

\[
\mathcal{G}\left(\boldsymbol{q},\boldsymbol{k},\boldsymbol{q}',\boldsymbol{k}'\right)=\frac{1}{2\pi}\iint g^{*}\left(\boldsymbol{q}-\frac{\boldsymbol{p}}{2},\boldsymbol{q}'-\frac{\boldsymbol{p}'}{2}\right)
\]

\begin{equation}
...\times g\left(\boldsymbol{q}+\frac{\boldsymbol{p}}{2},\boldsymbol{q}'+\frac{\boldsymbol{p}'}{2}\right)\exp\left\{ -i\left(\boldsymbol{k}\cdot\boldsymbol{p}-\boldsymbol{k}'\cdot\boldsymbol{p}'\right)\right\} d\boldsymbol{p}d\boldsymbol{p}'
\end{equation}

Together with Eq. (3) and (4), when we evaluate the WDF of the output
coherence function using Eq. (1), we obtain the WDF input-output relation.

\begin{equation}
\mathcal{W}\left(\boldsymbol{q},\boldsymbol{k}\right)=\iint\mathcal{G}\left(\boldsymbol{q},\boldsymbol{k},\boldsymbol{q}',\boldsymbol{k}'\right)\mathcal{W}\left(\boldsymbol{q}',\boldsymbol{k}'\right)d\boldsymbol{q}'d\boldsymbol{k}'
\end{equation}

where $\boldsymbol{q}$ and $\boldsymbol{k}$ are the space and spatial-frequency
variables, respectively. For simplicity, we have explicitly excluded
the temporal dependence. Also, In this letter, we have used the following
change of variables whenever applicable.

\begin{equation}
\begin{array}{ccc}
\mathrm{output} &  & \mathrm{input}\\
\boldsymbol{r}_{1}=\boldsymbol{q}-\boldsymbol{p}/2 & \quad\&\quad\quad & \boldsymbol{r}'_{1}=\boldsymbol{q}'-\boldsymbol{p}'/2\\
\boldsymbol{r}_{2}=\boldsymbol{q}+\boldsymbol{p}/2 &  & \boldsymbol{r}'_{2}=\boldsymbol{q}'+\boldsymbol{p}'/2
\end{array}
\end{equation}

\subsection{Incoherent light and Ensemble average}

\paragraph{Incoherent light source}

The results that we have obtained are generic, i.e., without any specific
assumption. However, in the coherence holography, it is required that
the output coherence function is wide-sense spatial stationary. This
stationarity condition is crucial for replacing the ensemble average
by the space average. So, if we assume that the source is wide-sense
stationary, i.e., $\Gamma\left(\boldsymbol{r}'_{1}+\boldsymbol{r}_{s},\boldsymbol{r}'_{2}+\boldsymbol{r}_{s}\right)=\Gamma\left(\boldsymbol{r}'_{1},\boldsymbol{r}'_{2}\right)$
and the green's function is shift-invariant, i.e., $g\left(\boldsymbol{r}'+\boldsymbol{r}_{s},\boldsymbol{r}+\boldsymbol{r}_{s}\right)=g\left(\boldsymbol{r}',\boldsymbol{r}\right)$
for a unit magnification $\left(\boldsymbol{r}_{s}=\boldsymbol{r}'_{s}\right)$.
The output coherence function becomes wide-sense stationary, verified
from the input-output relation for coherence function. But, in the
coherence holography, the source is obtained from illuminating the
hologram by spatially incoherent light. This is practically achieved
by passing the Laser light through rotating ground glass. The desired
ensemble-averaged degree of spatial coherence can be explicitly controlled
by several factors, given elsewhere \cite{asakura1970spatial}. The ensemble-averaged input coherence
function becomes $\Gamma\left(\boldsymbol{r}'_{1},\boldsymbol{r}'_{2}\right)=\mathrm{I}\left(\boldsymbol{r}'_{1}\right)\delta\left(\boldsymbol{r}'_{2}-\boldsymbol{r}'_{1}\right)$,
where $\mathrm{I}\left(\boldsymbol{r}'_{1}\right)$ is made proportional
to the intensity transmittance of the Hologram. It generally has a
complicated spatial fringe structure with limited spatial extent,
which means that such a source of practical interest is typically
non-stationary. Even though with a suitable propagation kernel the
output coherence function can still be made wide-sense stationary.
Two such green's functions are Fourier and Fresnel kernels \cite{takeda2005coherence,takeda2013spatial,takeda2014spatial}.

Now, for our analysis, we require an input WDF obtained from input
coherence function. So, if we put the above-defined coherence function
in Eq. (2), we get the following WDF,

\begin{equation}
\mathcal{W}\left(\boldsymbol{q}',\boldsymbol{k}'\right)=\mathrm{I}\left(\boldsymbol{q}'\right)
\end{equation}

As can be seen in Eq. $\left(7\right)$ that the input WDF is a function
of space variable only, which means the light radiates equally in
all directions (with $\mathrm{I}\left(\boldsymbol{q}'\right)\geq0$)
\cite{bastiaans2009wigner}.

\vspace{10pt}

\paragraph{Fourier kernel}

We consider a 2D Fourier kernel \cite{goodman2005introduction} which is defined as\newpage

\begin{equation}
g\left(\boldsymbol{r},\boldsymbol{r'}\right)=\frac{-ik_{o}}{2\pi f}\exp\left(ik_{o}f\right)\exp\left(-ik_{o}\frac{\boldsymbol{r}\cdot\boldsymbol{r}'}{f}\right)
\end{equation}

where $k_{\circ}$ is the mean wavenumber of the quasimonochromatic
light, and $f$ is the focal length of an aberration-free lens. Also,
the lens is assumed to be large enough so that the finite-aperture
effects can be neglected. The variables $\boldsymbol{r}$ and $\boldsymbol{r}'$
in the green's function have usual meaning. Now if we put Eq. $\left(8\right)$ in the \emph{double} WDF formula,
that is Eq. $\left(4\right)$, we obtain the following \emph{double}
WDF of the Fourier kernel
\begin{widetext}
\[
\mathcal{G}_{\mathrm{fourier}}\left(\boldsymbol{q},\boldsymbol{k},\boldsymbol{q}',\boldsymbol{k}'\right)=\frac{1}{2\pi}\iint\exp\left\{ i\frac{k_{o}}{f}\left(\boldsymbol{q}-\frac{\boldsymbol{p}}{2}\right)\cdot\left(\boldsymbol{q}'-\frac{\boldsymbol{p}'}{2}\right)\right\} \times\exp\left\{ -i\frac{k_{o}}{f}\left(\boldsymbol{q}+\frac{\boldsymbol{p}}{2}\right)\cdot\left(\boldsymbol{q}'+\frac{\boldsymbol{p}'}{2}\right)\right\} \mathrm{e}^{-i\left(\boldsymbol{k}\cdot\boldsymbol{p}-\boldsymbol{k}'\cdot\boldsymbol{p}'\right)}d\boldsymbol{p}d\boldsymbol{p}'
\]

\begin{equation}
=\frac{k_{o}}{2\pi f}\delta\left(\boldsymbol{q}'+\frac{f\boldsymbol{k}}{k_{o}}\right)\delta\left(\boldsymbol{k}'-\frac{k_{o}\boldsymbol{q}}{f}\right)
\end{equation}
\end{widetext}

Again, from Eq. $\left(7\right)$ and Eq. $\left(9\right)$, we obtain
the final output WDF using Eq. $\left(5\right)$

\begin{equation}
\mathcal{W}\left(\boldsymbol{q},\boldsymbol{k}\right)=\frac{k_{o}}{2\pi f}\mathrm{I}\left(-\frac{f\boldsymbol{k}}{k_{o}}\right)
\end{equation}
The output WDF in Eq. $\left(10\right)$, apart from a constant factor,
has the same intensity profile as that of the input WDF in Eq. $\left(5\right)$.
It is also a function of spatial-frequency only, which implies that
it has rotated in phase-space by $\pi/2$ radians in magnitude, and
the minus sign indicates the intensity profile flip. It should be
noted that the above formalism works well for the 1D case as well,
i.e., with scalar variables, when replacing the ensemble average by
the space average. However, the same isn't the case for the Fresnel
kernel mentioned below.

\paragraph{Fresnel kernel}

Here, we consider a 2-D Fresnel kernel. This formalism is only applicable
for the 2D case for a reason given elsewhere \citep{takeda2014spatial,takeda2013spatial} and will
also be evident when we cover the next section on space average. The
Fresnel kernel reads \cite{goodman2005introduction}
\begin{equation}
g\left(\boldsymbol{r},\boldsymbol{r}'\right)=\frac{-ik_{o}}{2\pi z}\mathrm{e}^{ik_{o}z}\exp\left\{ -i\frac{k_{o}}{2z}\left(\left|\boldsymbol{r}\right|^{2}-2\boldsymbol{r}\cdot\boldsymbol{r}'+\left|\boldsymbol{r}'\right|^{2}\right)\right\} 
\end{equation}

where $z$ is the distance between the source and the observation
planes, and the other symbols have their usual meanings. Now, putting
the Eq. $\left(11\right)$ in $\left(4\right)$ would yield the following
\emph{double }WDF of the Fresnel kernel
\begin{widetext}
\[
\mathcal{G}_{\mathrm{fresnel}}\left(\boldsymbol{q},\boldsymbol{k},\boldsymbol{q}',\boldsymbol{k}'\right)=\frac{1}{2\pi}\left(\frac{k_{o}}{2\pi z}\right)^{2}\iint\exp\left\{ i\frac{k_{o}}{2z}\left[\left|\boldsymbol{q}-\frac{\boldsymbol{p}}{2}\right|^{2}-2\left(\boldsymbol{q}-\frac{\boldsymbol{p}}{2}\right)\cdot\left(\boldsymbol{q}'-\frac{\boldsymbol{p}'}{2}\right)+\left|\boldsymbol{q}'-\frac{\boldsymbol{p}'}{2}\right|^{2}\right]\right\} 
\]

\[
...\times\exp\left\{ -i\frac{k_{o}}{2z}\left[\left|\boldsymbol{q}+\frac{\boldsymbol{p}}{2}\right|^{2}-2\left(\boldsymbol{q}+\frac{\boldsymbol{p}}{2}\right)\cdot\left(\boldsymbol{q}'+\frac{\boldsymbol{p}'}{2}\right)+\left|\boldsymbol{q}'+\frac{\boldsymbol{p}'}{2}\right|^{2}\right]\right\} \exp\left[-i\left(\boldsymbol{k}\cdot\boldsymbol{p}-\boldsymbol{k}'\cdot\boldsymbol{p}'\right)\right]d\boldsymbol{p}d\boldsymbol{p}'
\]

\begin{equation}
=\frac{k_{o}}{2\pi z}\delta\left(\boldsymbol{q}'-\left\{ \boldsymbol{q}+\frac{z\boldsymbol{k}}{k_{o}}\right\} \right)\delta\left(\boldsymbol{k}'-\frac{k_{o}}{z}\left\{ \boldsymbol{q}-\boldsymbol{q}'\right\} \right)
\end{equation}
\end{widetext}

and from the Eq. $\left(7\right)$ and Eq. $\left(12\right)$ the
output WDF using the Eq. $\left(5\right)$ becomes

\begin{equation}
\mathcal{W}\left(\boldsymbol{q},\boldsymbol{k}\right)=\frac{k_{o}}{2\pi z}\mathrm{I}\left(\boldsymbol{q}+\frac{z\boldsymbol{k}}{k_{o}}\right)
\end{equation}

Here also the output WDF, apart from a constant factor, has the same
intensity profile as given in Eq. $\left(7\right)$. Nevertheless,
the input WDF rotates in the phase-space such that the intensity profile
is defined on a hyperplane given by $\boldsymbol{q}+z\boldsymbol{k}/k_{\circ}=arbitrary\:constant$,
i.e., it rotates in the phase-space according to the value $z/k_{o}$
or how far the field propagates in fresnel regime.

\begin{figure*}[t]
\includegraphics[scale=0.56]{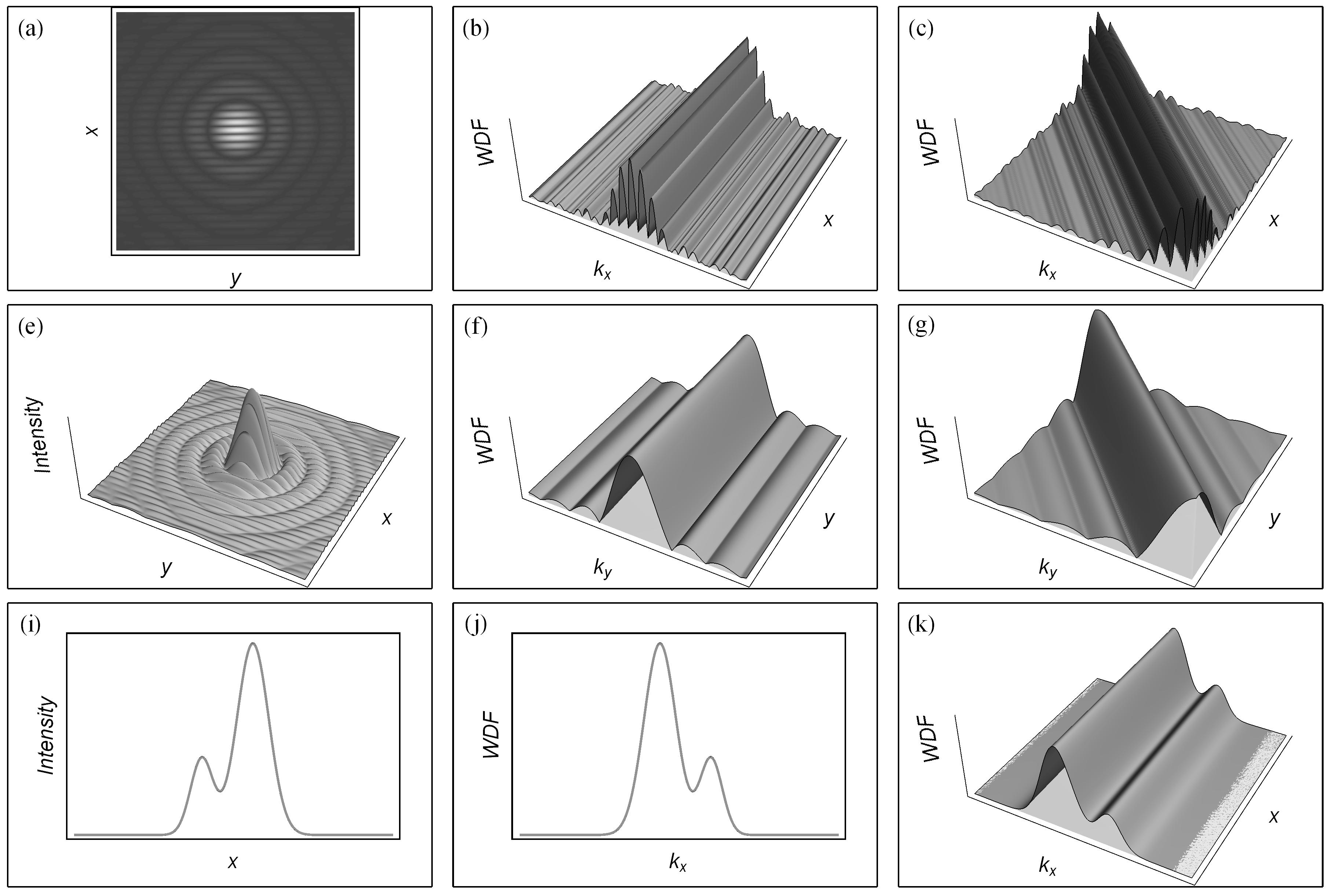}
\caption{\label{fig:wide} (a) \& (e) are respectively the 2D and 3D plots of the intensity transmittance
of the Fourier hologram illuminated by a plane wave
at some angle; (b) \& (f) are respectively the output WDFs for a particular
values of ($y,k_{y}$) and ($x,k_{x}$) in case of Fourier kernel;
(c) \& (g) are respectively the output WDFs for a particular values
of ($y,k_{y}$) and ($x,k_{x}$) in case of Fresnel kernel. Plots (i), (j), \& (k) are respectively the 1D intensity profile, vertical cross-section of its WDF along $k_{x}$-axis, and the entire WDF itself.}
\end{figure*}

\subsection{Space average}

\paragraph{Spatial coherence function \& WDF}

It is common to replace the ensemble average by the time average if
the statistical field is ergodic and stationary in time. However,
the same isn't the case in spatial statistical approach. In most practical
cases, neither the sources nor their diffraction fields are spatially
stationary. But, as mentioned earlier, the Fourier and Fresnel kernels
can produce wide-sense spatially stationary fields even though the
sources are spatially non-stationary. Then it is possible to replace
the ensemble average by the space average, which is defined differently
as per the kernels used, i.e., both Fourier and Fresnel kernels would
have different space averaging definitions. These results are well
established and given elsewhere \cite{takeda2013spatial,takeda2014spatial}. Here, we present the phase-space
interpretation of these replacements. Also, one may find the following
derivations in the paper mentioned above. Still, for consistency,
we have shown the steps explicitly and then implemented the WDF analogy
of those results.

Let us consider the definition of (output) spatial coherence function
\citep{takeda2014spatial,takeda2013spatial}

\begin{equation}
\Gamma^{s}\left(\boldsymbol{p}\right)=\int E^{*}\left(\boldsymbol{q}-\frac{\boldsymbol{p}}{2}\right)E\left(\boldsymbol{q}+\frac{\boldsymbol{p}}{2}\right)d\boldsymbol{q}
\end{equation}

where the field $E$ is given by Eq. $\left(1\right)$, the superscript
$s$ stands for the space average, and the variables are as per the
Eq. $\left(6\right)$. Here in Eq. $\left(14\right)$, the ensemble
average is replaced by the space average in the definition of coherence
function. This replacement is valid if the coherence function is wide-sense
stationary \cite{o2003introduction}. Correspondingly, the output WDF is given
by

\begin{equation}
\mathcal{W}^{s}\left(\boldsymbol{k}\right)=\int\Gamma^{s}\left(\boldsymbol{p}\right)\exp\left\{ -i\boldsymbol{k}\cdot\boldsymbol{p}\right\} d\boldsymbol{p}
\end{equation}

Next, we derive WDF of the spatial coherence functions for the Fourier
and Fresnel kernels.

\paragraph{Fourier kernel}

For a 2D Fourier kernel (Eq. $\left(8\right)$), if we use Eq. $\left(1\right)$
in Eq. $\left(14\right)$ and perform space averaging, we obtain

\[
\Gamma^{s}\left(\boldsymbol{p}\right)=\frac{k_{o}}{2\pi f}\iint\exp\left\{ -i\frac{k_{o}}{2f}\boldsymbol{p}\cdot(\boldsymbol{q}_{1}+\boldsymbol{q}_{2})\right\} \delta\left(\boldsymbol{q}_{1}-\boldsymbol{q}_{2}\right)
\]

\[
...\times E^{*}\left(\boldsymbol{q}_{1}\right)E\left(\boldsymbol{q}_{2}\right)d\boldsymbol{q}_{1}d\boldsymbol{q}_{2}
\]

\begin{equation}
=\frac{k_{o}}{2\pi f}\int\mathrm{I}\left(\boldsymbol{q}_{1}\right)\exp\left(-i\frac{k_{o}}{f}\boldsymbol{q}_{1}\cdot\boldsymbol{p}\right)d\boldsymbol{q}_{1}
\end{equation}

Now putting equation $\left(16\right)$ in equation $\left(15\right)$
yields

\begin{equation}
\mathcal{W}^{s}\left(\boldsymbol{k}\right)=\mathrm{I}\left(-\frac{f\boldsymbol{k}}{k_{o}}\right)
\end{equation}

where $\mathrm{I}\left(\boldsymbol{q}_{1}\right)=E^{*}\left(\boldsymbol{q}_{1}\right)E\left(\boldsymbol{q}_{1}\right)$
is the instantaneous intensity distribution over the source plane.
Eq. $(17)$ is the WDF of the output spatial coherence function obtained
from Fourier kernel. Also, note that the expression obtained in equation
$\left(17\right)$ is equivalent to equation $\left(10\right)$.

\paragraph{Fresnel kernel}

Again, in the case of 2D Fresnel kernel, we modify the spatial averaging
definition by considereing $\boldsymbol{q}=\boldsymbol{q}_{\Vert}+\boldsymbol{q}_{\bot}$,
where $\boldsymbol{q}_{\bot}$ and $\boldsymbol{q}_{\Vert}$ are the
orthogonal components of $\boldsymbol{q}$ such that $\boldsymbol{q}_{\bot}$
is perpendicular to $\boldsymbol{p}$ ($\boldsymbol{q}_{\bot}$$\cdot\boldsymbol{p}=0$).
Also we take space average with respect to $\boldsymbol{q}_{\bot}$
instead of $\boldsymbol{q'}$ in equation $\left(14\right)$ \cite{takeda2013spatial,takeda2014spatial},
i.e., 

\[
\Gamma^{s}\left(\boldsymbol{q}_{\Vert},\boldsymbol{p}\right)=\iiint g^{*}\left(\boldsymbol{q}_{\bot}+\boldsymbol{q}_{\Vert}-\frac{\boldsymbol{p}}{2},\boldsymbol{q}_{1}\right)E^{*}\left(\boldsymbol{q}_{1}\right)
\]

\begin{equation}
...\times g\left(\boldsymbol{q}_{\bot}+\boldsymbol{q}_{\Vert}+\frac{\boldsymbol{p}}{2},\boldsymbol{q}_{2}\right)E\left(\boldsymbol{q}_{2}\right)d\boldsymbol{q}_{1}d\boldsymbol{q}_{2}d\boldsymbol{q}_{\bot}
\end{equation}

putting Eq. $\left(11\right)$ for the green's function in Eq. $\left(18\right)$
and performing space average gives (after some algebraic steps)

\begin{widetext}
\[
\Gamma^{s}\left(\boldsymbol{q},\boldsymbol{p}\right)=\frac{k_{o}}{2\pi z}\exp\left(-i\frac{k_{o}}{z}\boldsymbol{p}\cdot\boldsymbol{q}_{\Vert}\right)\iint\delta\left(\boldsymbol{q}_{1}-\boldsymbol{q}_{2}\right)E^{*}\left(\boldsymbol{q}_{1}\right)E\left(\boldsymbol{q}_{2}\right)
\]

\[
...\times\exp\left[\frac{ik_{o}}{2z}\left\{ 2\boldsymbol{q}_{\Vert}\cdot(\boldsymbol{q}_{2}-\boldsymbol{q}_{1})+\boldsymbol{p}\cdot(\boldsymbol{q}_{2}+\boldsymbol{q}_{1})-(\boldsymbol{q}_{2}^{2}-\boldsymbol{q}_{1}^{2})\right\} \right]d\boldsymbol{q}_{1}d\boldsymbol{q}_{2}
\]

\begin{equation}
=\frac{k_{o}}{2\pi z}\exp\left(-i\frac{k_{o}}{z}\boldsymbol{p}\cdot\boldsymbol{q}\right)\int\exp\left(i\frac{k_{o}}{z}\boldsymbol{p}\cdot\boldsymbol{q}_{1}\right)\mathrm{I}\left(\boldsymbol{q}_{1}\right)d\boldsymbol{q}_{1}
\end{equation}

\end{widetext}

note that the substitution $\boldsymbol{p}\cdot\boldsymbol{q}_{\Vert}=\boldsymbol{p}\cdot\boldsymbol{q}$
is valid as $\boldsymbol{q}_{\bot}\cdot$$\boldsymbol{p}=0$. Now
putting Eq. $\left(19\right)$ in Eq. $\left(15\right)$ gives the
following output WDF of the modified spatial coherence function

\begin{equation}
\mathcal{W}^{s}\left(\boldsymbol{q},\boldsymbol{k}\right)=\mathrm{I}\left(\boldsymbol{q}+\frac{z\boldsymbol{k}}{k_{o}}\right)
\end{equation}

Here as well the obtained WDF is equivalent to the Eq. $(13)$. Hence,
one may get the output WDF from the spatial coherence functions instead
of the ensemble-averaged coherence function if the wide-sense stationarity
condition holds as discussed so far.

\section{Results and Discussion}

Here are the simulation results of the WDFs obtained for the Fourier
and Fresnel kernels. In Fig. $2$, (a) \& (e) represent the 2D and
3D plots of the intensity transmittance of the Fourier hologram of
a circular aperture illuminated by a plane wave at some angle to give
fringe pattern as shown. Usually, this Airy pattern is generated by
illuminating the hologram with the incoherent plane wave and then
taking the ensemble average. However, as we have seen that the output
coherence functions and the output WDFs are equivalent in the cases
of ensemble average and space average given the conditions are satisfied,
The Airy pattern here is simply generated by illuminating the hologram
with the plane wave once. It serves as the instantaneous intensity
profile for the spatial averaging approach. Also, it is to be noted
that the fringes occur along the x-axis. Next, the figures (b) and
(f) are the 2D WDFs obtained from the Fourier kernels. Although the
output WDF would be 4D, for the sake of interpreting its behaviour,
we have chosen a constant value of $(y,k_{y})$ in (b) where the normalized
WDF is one. In our case, it is the intermediate values of y and ky
of the image and its Fourier transform. As depicted from the Eq. $(10)$
and $(17)$ the input WDF rotates in the phase-space by $\pi/2$ radians
in magnitude to give the output WDF. The minus sign in Eq. (10) or
(17) represents an inversion of the intensify profile along k-axis.
However, since the intensity profile in our WDF expression is symmetric,
this behaviour is not apparent. Likewise, figure (f) shows the same
behaviour but for the constant value of $(x,k_{x})$ where the normalized
WDF value is one. Finally, figures (c) and (g) are the 2D WDFs obtained
from the Fresnel kernel. Here as well, the plots are shown for constant
values of $(y,k_{y})$ and $(x,k_{x})$ where the normalized WDF values
are one, respectively. In Eq. $(13)$ and $(20)$, it is shown that
the output WDF is the intensity profile with arguments as a linear
combination of space and spatial-frequency variables, i.e., $\boldsymbol{q}+z\boldsymbol{k}/k_{\circ}$.
It implies that the input WDF rotates in the phase-space according
to the value of $z/k_{\circ}$ or how far the coherence function propagates
in the Fresnel regime. In our case, we have simply taken this value
equal to unity, i.e., $z/k_{\circ}=1\:m^{2}$. Therefore, in
figures (c) and (g), the intensity profiles are defined on a hyperplane
$\boldsymbol{q}+\boldsymbol{k}=arbitrary\:constant$. In other words,
the input WDF rotates in phase-space by $\pi/4$ radians in magnitude. As mentioned earlier, the formalism for the Fourier kernel is also valid in 1D case. Figure (i) shows a 1D intensity profile obtained from the superposition of two Gaussian profiles with different amplitudes and shifted means. Also, the smaller amplitude Gaussian profile is to the left of the bigger one. In figures (j) and (k), the corresponding vertical cross-section of the WDF along $k_{x}$-axis and the entire WDF are shown, respectively. One may notice that apart from a $\pi/2$ rotation in phase-space, the intensity profiles in figures (j) and (k) are inverted, demonstrating the effect of minus sign in the output WDF (Eq. $(10)$ and $(17)$) obtained from the Fourier kernel.

\section{Conclusion}s
In this paper, we showed, by suplimenting the unlying theory of coherence
holography, that in phase-space the output WDFs obtained from the
ensemble-averaged and space-averaged cohererence functions are equivalent
to each other provided the source in spatially incoherent and the
propagation kernel is either Fourier or Fresnel. Also, in the case
of Fresnel kernel, the space average definition is modified accordingly.
We also showed the behaviours of these output WDFs in the phase-space.
For the Fourier kernel, the WDF rotates in phase-space by $\pi/2$
radians in magnitude and the intensity profile is inverted; whereas,
for the Fresnel kernel, the WDF rotates in phase-space depending on
how far the coherence function propagates in the Fresnel regime, i.e.,
$z/k_{\circ}$ value. These results are in general valid for any optical
phenomena where these conditions are statisfied. Hence, one may replace
the output WDF obtained from ensemble-averaged coherence function
by the WDF obtained from the space-averaged coherence function.

\begin{acknowledgments}
Part of this work was carried out under SERB project no. CRG/2019/000026, India. Also, Rishabh K. B. carried out this work as a part of his master thesis.
\end{acknowledgments}

\bibliography{apssamp}% Produces the bibliography via BibTeX.

\providecommand{\noopsort}[1]{}\providecommand{\singleletter}[1]{#1}%
\begin{thebibliography}{23}
\providecommand{\natexlab}[1]{#1}
\providecommand{\url}[1]{\texttt{#1}}
\expandafter\ifx\csname urlstyle\endcsname\relax
  \providecommand{\doi}[1]{doi: #1}\else
  \providecommand{\doi}{doi: \begingroup \urlstyle{rm}\Url}\fi

\bibitem[Asakura(1970)]{asakura1970spatial}
Toshimitsu Asakura.
\newblock Spatial coherence of laser light passed through rotating ground
  glass.
\newblock \emph{Opto-electronics}, 2\penalty0 (3):\penalty0 115--123, 1970.

\bibitem[Bastiaans(1978)]{bastiaans1978wigner}
Martin~J Bastiaans.
\newblock The wigner distribution function applied to optical signals and
  systems.
\newblock \emph{Optics communications}, 25\penalty0 (1):\penalty0 26--30, 1978.

\bibitem[Bastiaans(1980)]{bastiaans1980wigner}
Martin~J Bastiaans.
\newblock The wigner distribution function and its applications to optics.
\newblock In \emph{AIP conference proceedings}, volume~65, pages 292--312.
  American Institute of Physics, 1980.

\bibitem[Bastiaans et~al.(2009)]{bastiaans2009wigner}
Martin~J Bastiaans et~al.
\newblock Wigner distribution in optics, 2009.

\bibitem[Bastiaans(1979{\natexlab{a}})]{bastiaans1979transport}
MJ~Bastiaans.
\newblock Transport equations for the wigner distribution function.
\newblock \emph{Optica Acta: International Journal of Optics}, 26\penalty0
  (10):\penalty0 1265--1272, 1979{\natexlab{a}}.

\bibitem[Bastiaans(1979{\natexlab{b}})]{bastiaans1979wigner}
MJ~Bastiaans.
\newblock Wigner distribution function and its application to first-order
  optics.
\newblock \emph{JOSA}, 69\penalty0 (12):\penalty0 1710--1716,
  1979{\natexlab{b}}.

\bibitem[City(1948)]{ville2orie}
J~City.
\newblock Theory and applications of the notion of analytical signal.
\newblock \emph{CeT, Telecommunications Laboratory of the Société Alsacienne
  de Construction Mecanique}, 2, 1948.

\bibitem[Gabor(1948)]{gabor1948new}
Dennis Gabor.
\newblock A new microscopic principle, 1948.

\bibitem[Goodman(2005)]{goodman2005introduction}
Joseph~W Goodman.
\newblock \emph{Introduction to Fourier optics}.
\newblock Roberts and Company Publishers, 2005.

\bibitem[Khan et~al.(2011)Khan, Taj, Jaffri, and Ijaz]{khan2011cross}
Nabeel~Ali Khan, Imtiaz~Ahmad Taj, M~Noman Jaffri, and Salman Ijaz.
\newblock Cross-term elimination in wigner distribution based on 2d signal
  processing techniques.
\newblock \emph{Signal Processing}, 91\penalty0 (3):\penalty0 590--599, 2011.

\bibitem[Kim et~al.(2008)Kim, Min, Lee, and Poon]{kim2008optical}
Hwi Kim, Sung-Wook Min, Byoungho Lee, and Ting-Chung Poon.
\newblock Optical sectioning for optical scanning holography using phase-space
  filtering with wigner distribution functions.
\newblock \emph{Applied optics}, 47\penalty0 (19):\penalty0 D164--D175, 2008.

\bibitem[Leith and Upatnieks(1962)]{leith1962reconstructed}
Emmett~N Leith and Juris Upatnieks.
\newblock Reconstructed wavefronts and communication theory.
\newblock \emph{JOSA}, 52\penalty0 (10):\penalty0 1123--1130, 1962.

\bibitem[Leith and Upatnieks(1963)]{leith1963wavefront}
Emmett~N Leith and Juris Upatnieks.
\newblock Wavefront reconstruction with continuous-tone objects.
\newblock \emph{JOSA}, 53\penalty0 (12):\penalty0 1377--1381, 1963.

\bibitem[Lohmann et~al.(2002)Lohmann, Testorf, and
  Ojeda-Castaneda]{lohmann2002holography}
Adolf~W Lohmann, Markus~E Testorf, and Jorge Ojeda-Castaneda.
\newblock Holography and the wigner function.
\newblock In \emph{Holography: A Tribute to Yuri Denisyuk and Emmett Leith},
  volume 4737, pages 77--88. International Society for Optics and Photonics,
  2002.

\bibitem[Oh and Barbastathis(2009)]{oh2009wigner}
Se~Baek Oh and George Barbastathis.
\newblock Wigner distribution function of volume holograms.
\newblock \emph{Optics letters}, 34\penalty0 (17):\penalty0 2584--2586, 2009.

\bibitem[O'Neill(2003)]{o2003introduction}
Edward~L O'Neill.
\newblock \emph{Introduction to statistical optics}.
\newblock Courier Corporation, 2003.

\bibitem[Qian and Chen(1999)]{qian1999joint}
Shie Qian and Dapang Chen.
\newblock Joint time-frequency analysis.
\newblock \emph{IEEE Signal Processing Magazine}, 16\penalty0 (2):\penalty0
  52--67, 1999.

\bibitem[Situ and Sheridan(2007)]{situ2007holography}
Guohai Situ and John~T Sheridan.
\newblock Holography: an interpretation from the phase-space point of view.
\newblock \emph{Optics letters}, 32\penalty0 (24):\penalty0 3492--3494, 2007.

\bibitem[Takeda(2013)]{takeda2013spatial}
Mitsuo Takeda.
\newblock Spatial stationarity of statistical optical fields for coherence
  holography and photon correlation holography.
\newblock \emph{Optics letters}, 38\penalty0 (17):\penalty0 3452--3455, 2013.

\bibitem[Takeda et~al.(2005)Takeda, Wang, Duan, and
  Miyamoto]{takeda2005coherence}
Mitsuo Takeda, Wei Wang, Zhihui Duan, and Yoko Miyamoto.
\newblock Coherence holography.
\newblock \emph{Optics express}, 13\penalty0 (23):\penalty0 9629--9635, 2005.

\bibitem[Takeda et~al.(2014)Takeda, Wang, Naik, and Singh]{takeda2014spatial}
Mitsuo Takeda, Wei Wang, Dinesh~N Naik, and Rakesh~K Singh.
\newblock Spatial statistical optics and spatial correlation holography: a
  review.
\newblock \emph{Optical Review}, 21\penalty0 (6):\penalty0 849--861, 2014.

\bibitem[Torre(2005)]{torre2005linear}
Amalia Torre.
\newblock \emph{Linear ray and wave optics in phase space: bridging ray and
  wave optics via the Wigner phase-space picture}.
\newblock Elsevier, 2005.

\bibitem[Wigner(1932)]{wigner1932phys}
E~Wigner.
\newblock Phys. rev.
\newblock \emph{On the Quantum Correction for Thermodynamic Equilibrium},
  40:\penalty0 pp--749, 1932.

\end{thebibliography}

\end{document}